\documentclass[prc,aps,floatfix,groupedaddress,amsmath,amssymb,twocolumn]{revtex4}

\usepackage{graphicx}% Include figure files
\usepackage{dcolumn}%Align table columns on decimal point
\usepackage{bm}% bold math
\usepackage{color}

\usepackage[section]{placeins}
\usepackage{graphicx,epsfig,accents}
\usepackage{amsmath}
\usepackage{amsfonts}
\usepackage{braket}
\usepackage{fancyhdr}
\usepackage{hyperref}
\usepackage[utf8]{inputenc}
\usepackage[T1]{fontenc}
\usepackage{amsthm}
\usepackage{hhline}
\usepackage{multirow}
\usepackage[figuresleft]{rotating}
\usepackage{gensymb}
\usepackage{xcolor}
\def\beq{\begin{equation}}
\def\eeq{\end{equation}}
\def\beqa{\begin{eqnarray}}
\def\eeqa{\end{eqnarray}}

\makeatletter
\DeclareRobustCommand{\cev}[1]{%
  \mathpalette\do@cev{#1}%
}
\newcommand{\do@cev}[2]{%
  \fix@cev{#1}{+}%
  \reflectbox{$\m@th#1\vec{\reflectbox{$\fix@cev{#1}{-}\m@th#1#2\fix@cev{#1}{+}$}}$}%
  \fix@cev{#1}{-}%
}
\newcommand{\fix@cev}[2]{%
  \ifx#1\displaystyle
    \mkern#23mu
  \else
    \ifx#1\textstyle
      \mkern#23mu
    \else
      \ifx#1\scriptstyle
        \mkern#22mu
      \else
        \mkern#22mu
      \fi
    \fi
  \fi
}

\makeatother

\newcommand{\up}{\uparrow}
\newcommand{\dw}{\downarrow}
\newcommand{\en}{\varepsilon_{n\sigma}(p)}
\newcommand{\ep}{\varepsilon_{p\sigma}(p_z,N_p)}
\newcommand{\ee}{\varepsilon_{e\sigma}(p_z,N_e)}

\begin{document}

%%%%%%%%%%%%%%%%%%%%%%%%%%%%%%%%%%%%%%%%%%%%%%%%%%%%%%%%%%%%%%%%%%%%%%%%%%%%%%%%%%%%

%%%

\title{Hot and highly magnetized neutron star matter properties with Skyrme interactions}

\author{Omar G. Benvenuto$^{1,2\footnote{Member of the Carrera del Investigador Científico, Comisión de Investigaciones Científicas de la Provincia de Buenos Aires, La Plata, Argentina.}}$, Eduardo Bauer$^{1,3}$   and Isaac Vida\~na$^4$}
\address{$^1$Facultad de Ciencias Astron\'onicas y Geof\'{i}sicas, Universidad Nacional de la Plata, Paseo del Bosque S/N, B1900FWA La Plata, Argentina
}
\address{$^2$Instituto de Astrof\'{i}sica de la Plata, CCT-CONICET-UNLP, Argentina}
\address{$^3$IFLP, CCT-La Plata CONICET, Argentina}
\address{$^4$Istituto Nazionale di Fisica Nucleare, Sezione di Catania, Dipartimento di Fisica e Astronomia ``Ettore Majorana'', Universit\`a di Catania, Via Santa Sofia 64, I-95123 Catania, Italy}

%%%

\begin{abstract}

We study the properties of  hot and dense neutron star matter under the presence of strong magnetic fields using two Skyrme interactions, namely the LNS and the BSk21 ones. Asking for $\beta$--stability and charge neutrality, we construct the equation of state of the system and analyze its composition for a range of densities, temperatures and magnetic field intensities of interest for the study of supernova and proto-neutron star matter, with a particular interest on the degree of spin-polarization of the different components. The results show that system configurations with larger fractions of spin up protons and spin down neutrons and electrons are energetically favored over those with larger fractions of spin down protons and spin up neutrons and electrons. The effective mass of neutrons and protons is found to be in general larger for the more abundant of their spin projection component, respectively, spin down neutrons and spin up protons. The effect of the magnetic field on the Helmhotz total free energy density, pressure and isothermal compressibility of the system is almost negligible for all the values of the magnetic field considered.

\end{abstract}

%%%

\maketitle

%%%%%%%%%%%%%%%%%%%%%%%%%%%%%%%%%%%%%%%%%%%%%%%%%%%%%%%%%%%%%%%%%%%

\section{Introduction}
\label{sec:intro}

The properties of neutron stars \cite{PharosBook} can be drastically modified due to the presence of strong magnetic fields. Most radio pulsars and accreting neutron stars in X-ray binaries present surface magnetic fields with intensities in the range $10^{12}-10^{13}$ G \cite{ShapiroTeukolsky}. Also recycled millisecond pulsars and old neutron stars in low-mass X-ray binaries have high surface fields of about $10^{8}-10^{9}$ G \cite{Lewin95,Lyne98}. In the surface of soft-gamma-ray repeaters and slowly spinning anomalous X-ray pulsars, the so-called ``magnetars'', the field can reach values of the order of $10^{14}-10^{15}$ G \cite{Duncan92,Paczynski92,Thompson95,Thompson96}. The intensity of these fields may grow by several orders of magnitude in the dense interior of all these compact objects
up to an upper limit, $B\leq10^{18}(M/1.4M_\odot)(R/10\,\mbox{km})^{-2}$ G, which follows from the virial theorem of magnetohydrostatic equilibrium
\cite{ShapiroTeukolsky,Chandrasekhar53}.
However, the origin of such large intensities remains still uncertain. These strong fields could be the fossil remnants from those of the progenitor star \cite{Tatsumi06}, or alternatively, they could be generated after the formation of the neutron star by some kind of dynamo process due to some long-lived electric currents flowing in the highly conductive neutron star material \cite{Thompson93}. Another possibility that has long been considered by many authors, with however contradictory results, is that these fields result from a spontaneous phase transition to a ferromagnetic state at densities corresponding to theoretically stable neutron stars (see, {\it e.g.,}
Refs.\ \cite{Brownell69,Rice69,Clark69,Silvertein69,Clark69b,Ostgaard70,Pearson70,Pandharipande72,Backman73,Haensel75,Jackson82,Vidaurre84,Kutschera89,Marcos91,Kutschera94,Bernardos96,Fantoni01,Vidana02,Vidana02b,Rios05,Lopez06,Bombaci06,Sammarruca07,Perez09,Perez09b,Bigdeli10,Sammarruca11}). Whatever is the origin of these fields, however, it is clear that the study of nuclear matter under the influence of strong magnetic fields is fundamental for a complete understanding of the magnetic properties of neutron stars.

Several authors have studied the magnetization of symmetric nuclear matter and pure neutron matter \cite{Khalilov02,PerezGarcia08,Aguirre11,PerezGarcia11,Aguirre13,Aguirre14,Aguirre15}. The magnetization of $\beta$-stable neutron star matter, however, has received less attention in the literature. Blandford and Hernquist in Ref.\ \cite{Blandford82}, for instance,
studied extensively the magnetization of $\beta$-stable matter for a single-component electron gas and for the crust matter of neutron stars. This study was generalized years later by Broderick {\it et al.} in Ref.\ \cite{Broderick00} by including also the contribution of neutrons and protons. The effect of the density dependence of the nuclear symmetry energy  on the magnetization of $\beta$-stable matter was studied few years ago by Dong {\it et al.} \cite{Dong13}, concluding that the magnetic susceptibility of protons, electrons and muons can be larger than that of neutrons. These authors found also that the anomalous magnetic moment of protons enhances their magnetic susceptibility to the point that it can be one of the main contributions and, therefore, should not be neglected. In Ref.\ \cite{Rabhi15} Rabhi {\it et al.} employed a relativistic mean field (RMF) approach to study
the effects of strong magnetic field on the proton and neutron spin polarization, and the magnetic susceptibility in asymmetric nuclear matter. The authors of this work showed
that magnetic fields of about $10^{16}-10^{17}$ G have noticeable effects on the range of densities of interest for the study of neutron star crusts. They found also that, although the magnetic susceptibility of protons is lager for weaker fields, the one of neutrons becomes of the same order or even larger at subsaturation densities for small values of the proton fraction when the fields are larger than $\sim 10^{16}$ G. In the recent years, the Coimbra group has published several works devoted to the study the effects of strong magnetic fields on the crust-core transition and the inner crust of neutron stars
\cite{Fang16,Fang17,Fang17b,Avancini18,Sengo20,Pais21,Ferreira21}. In these works the Vlasov equation is used to determine, using a RMF approach, the dynamical spinodal instability region which gives a good estimation of the crust-core transition in neutron stars. The results of these works show that strong magnetic fields of the order of $10^{15}-10^{17}$ G have a large effect on the spinodal region, defining the crust-core transition as a succession of stable and unstable regions due to the opening of new Landau levels. The results of these studies show also that sufficiently strong magnetic fields can significantly modify the extension of the unstable region and, therefore, \ the crust of magnetized neutron stars. The effect of temperature on the crust-core transition of magnetar was studied by this group in Ref.\ \cite{Fang17b} and very recently also in Ref.\ \cite{Ferreira21}. In these works, the authors showed that  the effect on the extension of the crust-core transition is washed away for temperatures above $10^9$ K for magnetic field intensities $\sim 5\times 10^{16}$ G but may still persist if a magnetic field as high as $\sim 5\times 10^{17}$ G is considered. They found that for lower temperatures, the effect of the magnetic field on the crust-cross transition is noticeable and grows as the temperature decreases.

In this work we study the properties of an electrically neutral system of neutrons, protons and electrons in equilibrium with respect to the weak interaction ($\beta$-equilibrium), at finite temperature and in the presence of  strong magnetic fields using Skyrme interactions. We consider a range of densities, temperatures and magnetic field intensities of interest for the study of supernova and proto-neutron star matter. We construct the equation of state (EoS) of the system and analyze its composition, with a particular interest on the degree of spin-polarization of the different components. The isothermal compressibility is also calculated and analyzed. The final scope of this work is to establish a framework for a future study of neutrino propagation in hot dense neutron star matter under the presence of strong magnetic fields. Neutrino cross sections and, consequently, the neutrino mean free path can be substantially affected by the presence of strong magnetic fields in neutron stars. For instance, the emission of neutrinos becomes asymmetric. It depends on the direction of the neutrino, as it was recently shown in Ref.\ \cite{TorresPatino19} where the neutrino mean free path in hot pure neutron matter under the presence of strong magnetic fields was analyzed by two of the authors of the present work. The extension of this analysis for the propagation of neutrinos in (more realistic) $\beta$-stable matter is left for the near future, and it will be based on the results of the present work.

The paper is organized in the following way.  The formalism employed to determine the properties of hot neutron star matter under the presence of a external magnetic field is described in Sec.\ \ref{sec:formalism}. Results are presented and discussed in Sec.\ \ref{sec:res}. Finally, a short summary and the main conclusions of this work are given in Sec.\ \ref{sec:suc}.

%%%%%%%%%%%%%%%%%%%%%%%%%%%%%%%%%%%%%%%%%%%%%%%%%%%%%%%%%%%%%%%%%%%
\section{Formalism}
\label{sec:formalism}

As said in the introduction, in this work we consider an electrically neutral system of neutrons, protons and electrons in $\beta$-equilibrium, at finite temperature and in the presence of strong magnetic fields. The physical state of the system can be obtained by minimizing a function $F$ which is constructed from the Helmhotz total free energy density ${\cal F}$ of the system and two constraints that express, respectively, the conditions of baryon number conservation and electrical charge neutrality
\begin{equation}
F={\cal F}+\alpha\left(\rho-\sum_{\tau=n,p}\sum_{\sigma=\up,\dw}\rho_{\tau\sigma}\right)
+\beta\sum_{\sigma=\up,\dw}\left(\rho_{p\sigma}-\rho_{e\sigma}
\right) \ .
\label{eq:functional}
\end{equation}
Here $\rho$ is the total baryon number density, and $\rho_{\tau\sigma}$ and $\rho_{e\sigma}$ are, respectively, the densities of neutrons ($\tau=n$), protons ($\tau=p$) and electrons ($e$), with spin up ($\sigma=\,\up$) or spin down ($\sigma=\,\dw$) projections. The quantities $\alpha$ and $\beta$ are the two Lagrange multipliers associated to each one of the constraints. The minimization of $F$ requires its partial derivatives with respect to the particle densities and the two multipliers to be zero, {\it i.e.,}
\begin{eqnarray}
\frac{\partial F}{\partial \rho_{\tau\sigma}}&=&0\ , \,\, \frac{\partial F}{\partial \rho_{e\sigma}}=0 \ , \nonumber \\
\frac{\partial F}{\partial \alpha}&=&0\ , \,\,\,\,\,\, \frac{\partial F}{\partial \beta}=0 \ .
\label{eq:mini1}
\end{eqnarray}

Remembering that the chemical potential of a particle $i$ is $\mu_i=\partial {\cal F}/\partial \rho_i$, the above conditions yield the following set of eight equations
\begin{equation}
\mu_i -b_i\alpha+q_i\beta =0, \,\,\, i=n_\up, n_\dw, p_\up, p_\dw, e_\up, e_\dw \ ,
\label{chempot}
\end{equation}
\begin{equation}
\rho=\sum_{\tau=n,p}\sum_{\sigma=\up,\dw}\rho_{\tau\sigma} \ ,
\label{baryon}
\end{equation}
\begin{equation}
\sum_{\sigma=\up,\dw}\rho_{p\sigma}=\sum_{\sigma=\up,\dw}\rho_{e\sigma} \ ,
\label{charge}
\end{equation}
where $b_i$ is the baryon number of particle $i$ and $q_i$ its electric charge. Eliminating the Lagrange multipliers $\alpha$ and $\beta$, one can obtain a set of relations among the chemical potentials of the different particles. In general there are as many independent chemical potentials as there are conserved charges, and all the others can be written in terms of them. In the case of neutron stars there are only two conserved charges (baryon number and electric charge), and we chose $\mu_{n_\up}$ and $\mu_{e_\dw}$ as the two independent chemical potentials associated with them. Applying now Eq.\ (\ref{chempot}) to $n_\up$ and $e_\up$ one finds
\begin{equation}
\alpha = \mu_{n_\up}\ , \,\,\, \beta=\mu_{e_\up} \ .
\end{equation}
Therefore, we can write
\begin{equation}
\mu_i = b_i\mu_{n_\up}-q_i\mu_{e_\up}\ , \,\,\, i=n_\up, n_\dw, p_\up, p_\dw, e_\up, e_\dw  \ ,
\label{chempot2}
\end{equation}
which together with Eqs.\ (\ref{baryon}) and (\ref{charge}) allow to determine the composition of the system. Note that
\begin{equation}
\mu_{n_\up}=\mu_{n_\dw}\ , \,\,\,\, \mu_{p_\up}=\mu_{p_\dw}\ , \,\,\,\, \mu_{e_\up}=\mu_{e_\dw}\ ,
\label{chempot3}
\end{equation}
that is, in the physical state the chemical potential of each species is independent of their spin projection.

We consider the interaction of electrons only with the magnetic field, but not between electrons with themselves or with protons, whereas to describe the in-medium interactions among the nucleons we employ Skyrme forces. In particular, we use the LNS interaction developed by Cao {\it et al.,} \cite{Cao06} and the interaction BSk21 \cite{Goriely10} of the Brussels--Montreal group. We note that the BSk21 interaction contains two new terms, in addition to the usual ones of the Skyrme force, that are introduced in order to avoid the appearance of ferromagnetic instability
at high densities, a general feature of all the conventional Skyrme forces developed in the past, as it is the case of the LNS one.

The total energy density of the system is given by
\begin{equation}
{\cal E}={ \cal E}_{nucl}+{\cal E}_{elec}+\mu_NB\left(2L_p+\rho_p-\frac{g_p}{2}W_p-\frac{g_n}{2}W_n\right) \ ,
\label{eq:energden}
\end{equation}
where ${\cal E}_{nucl}$ is the nuclear contribution, obtained in our case using the Hartree--Fock approximation with the Skyrme interaction, ${\cal E}_{elec}$ is the electron one, and
the last term shows the explicit dependence of the energy density on the magnetic field.
In the next we present separately these three contributions. We note that throughout all this work we use natural units in which $\hbar=c=1$.

The nuclear contribution can be written in a compact form as
\begin{widetext}
\begin{equation}
{\cal E}_{nucl}=\sum_{\tau=n,p}\sum_{\sigma=\up,\dw}\frac{K_{\tau \sigma}}{2m^*_{\tau \sigma}}
+\frac{1}{16}\left[
(a_0+a_2w^2)\rho^2
+a_1\left(\sum_{\tau=n,p}W_\tau\right)^2
+a_3\left(\sum_{\tau=n,p}I_\tau W_\tau\right)^2
\right] \ .
\label{eq:energden_nuc}
\end{equation}
\end{widetext}
Here
\begin{equation}
w=\frac{1}{\rho}\left[\sum_{\sigma=\pm 1}\rho_{n\sigma}-\sum_{\sigma=\pm 1}\rho_{p\sigma}
\right]
\label{eq:eq:totden}
\end{equation}
is the isospin asymmetry, with
\begin{equation}
\rho_{n\sigma}=\frac{1}{(2\pi)^3}\int d^3\vec p f_{n\sigma}(\en,T)
\label{eq:rhon}
\end{equation}
and
\begin{equation}
\rho_{p\sigma}=\frac{eB}{(2\pi)^2}\sum_{N_p}\int_{-\infty}^\infty dp_z f_{p\sigma}(\ep,T)
\label{eq:rhop}
\end{equation}
being the partial densities of neutron and protons with spin projection $\sigma$. The effective mass of a spin up or down nucleon is given by
\begin{eqnarray}
m^*_{\tau\sigma}&=&\Big[
\frac{1}{m_\tau}+\frac{1}{4}(b_0+c_0-(b_2+c_2)wI_\tau)\rho \nonumber \\
&+&\frac{s_\sigma}{4}\sum_{\tau'=n,p}(b_1+c_1+(b_3+c_3)I_\tau I_{\tau'})W_{\tau'}
\Big]^{-1} \ ,
\label{eq:effmass}
\end{eqnarray}
where $m_\tau$ is the bare mass of the nucleon, $I_\tau=1(-1)$ for protons (neutrons), and $s_\sigma=1(-1)$ if the spin projection is up (down). The coefficients $a_0,\cdot\cdot\cdot, a_3$, $b_0,\cdot\cdot\cdot, b_3$ and $c_0,\cdot\cdot\cdot, c_3$ are given in terms of  the parameters $t_{i=0,\cdot\cdot\cdot, 5}, x_{i=0,\cdot\cdot\cdot, 5}$, $\gamma$ and  $\beta$ of the LNS and BSk21 interactions through the relations
\begin{eqnarray}
a_0&=&6t_0+t_3\rho^\gamma \nonumber \\
a_1&=&-2t_0(1-2x_0)-\frac{t_3}{3}(1-2x_3)\rho^\gamma \nonumber \\
a_2&=&-2t_0(1+2x_0)-\frac{t_3}{3}(1+2x_3)\rho^\gamma \nonumber \\
a_3&=&-2t_0-\frac{t_3}{3}\rho^\gamma \nonumber \\
b_0&=&\frac{1}{2}[3t_1+t_2(5+4x_2)] \nonumber \\
b_1&=&\frac{1}{2}[t_2(1+2x_2)-t_1(1-2x_1)] \nonumber \\
b_2&=&\frac{1}{2}[t_2(1+2x_2)-t_1(1+2x_1)] \nonumber \\
b_3&=&\frac{1}{2}(t_2-t_1) \nonumber \\
c_0&=&\frac{1}{2}[3t_4\rho^\beta+t_5\rho^\gamma(5+4x_5)] \nonumber \\
c_1&=&\frac{1}{2}[t_5\rho^\gamma(1+2x_5)-t_4\rho^\beta(1-2x_4)] \nonumber \\
c_2&=&\frac{1}{2}[t_5\rho^\gamma(1+2x_5)-t_4\rho^\beta(1+2x_4)] \nonumber \\
c_3&=&\frac{1}{2}(t_5\rho^\gamma-t_4\rho^\beta) \ .
\label{eq:paramet}
\end{eqnarray}
We note that the terms with the parameters $t_4, t_5, x_4, x_5$ and $\beta$ are absent in the case of the LNS force and, therefore, the coefficients $c_0, \cdot\cdot\cdot, c_3$ are set equal to zero in this case. The quantities $K_{\tau\sigma}$ and $W_\tau$ appearing in Eqs.\ (\ref{eq:energden}), (\ref{eq:energden_nuc}) and (\ref{eq:effmass}) are related, respectively, with the kinetic energy and the spin asymmetry density, and are defined as
\begin{equation}
K_{n\sigma}=\frac{1}{(2\pi)^3}\int d^3\vec p p^2f_{n\sigma}(\en,T)
\label{eq:kn}
\end{equation}
\begin{equation}
W_{n}=\rho_{n\up}-\rho_{n\dw}
\label{eq:wn}
\end{equation}
in the case of neutrons, and
\begin{equation}
K_{p\sigma}=\frac{eB}{(2\pi)^2}\sum_{N_p}\int_{-\infty}^\infty dp_z p_z^2 f_{p\sigma}(\ep,T)
\label{eq:kp}
\end{equation}
\begin{equation}
W_{p}=\rho_{p\up}-\rho_{p\dw}
\label{eq:wp}
\end{equation}
for protons. Note that  when the magnetic field is assumed to be along the direction of the z-axis, due to the Landau quantization, the $x$ and $y$ components of the proton momentum spread over a bound region of area $2\pi eB$ in the $p_x-p_y$ plane, whereas the $z$ one is not bound and varies continuously. Therefore, the contribution of the protons to
any macroscopic quantity per unit volume is evaluated by means of the replacement $\int d^3\vec p/(2\pi)^3\rightarrow eB\int dp_z/(2\pi)^2 \sum_{N_p}$. The sums over the proton Landau levels $N_p$, in Eqs.\ (\ref{eq:kp}) and (\ref{eq:wp}), run from 0 up to a maximum level which is determined numerically, as it is explained with detail in the appendix B of Ref.\ \cite{Bauer20}.

The functions $f_{n\sigma}(\en,T)$ and $f_{p\sigma}(\ep,T)$ in Eqs.\ (\ref{eq:rhon})-(\ref{eq:rhop}) and (\ref{eq:kn})-(\ref{eq:wp}) are the corresponding neutron and proton Fermi--Dirac momentum distributions
\begin{equation}
f_{\tau\sigma}(\varepsilon_{\tau\sigma},T)=\left[1+\mbox{exp}\left(\frac{\varepsilon_{\tau\sigma}-\mu_{\tau\sigma}}{T}\right) \right]^{-1} \ ,
\label{eq:FD}
\end{equation}
where the neutron and proton single-particle energies are, respectively
\begin{equation}
\varepsilon_{n\sigma}(p)=m_n+\frac{p^2}{2m^*_{n\sigma}}+\frac{1}{8}v_{n\sigma}-\mu_Ng_n\frac{s_\sigma}{2} B \ ,
\label{eq:enern}
\end{equation}
and
\begin{eqnarray}
\varepsilon_{p\sigma}(p_z,N_p)&=&m_p+\frac{p^2_z}{2m^*_{p\sigma}}+\frac{1}{8}v_{p\sigma} \nonumber \\
&+&\mu_NB\left(2N_p+1-g_p\frac{s_\sigma}{2}\right) \ .
\label{eq:enerp}
\end{eqnarray}
Here $\mu_N=3.15245 \times 10^{-18}$ MeV/G is the nuclear magneton, $g_n=-3.826$ and $g_p=5.586$ are, respectively, the neutron and proton g-factors which take into account their anomalous magnetic moments, and
\begin{eqnarray}
v_{\tau\sigma}&=&(a_0 - a_2 w I_\tau)\, \rho
+ s_\sigma \sum_{\tau'=n,p} (a_1 + a_3 I_\tau I_{\tau'}) W_{\tau'}
\nonumber \\
&+& \sum_{\tau'=n,p} \sum_{\sigma'=\up,\dw}(b_0+c_0 +(b_2+c_2)I_\tau I_{\tau'})
K_{\tau'\sigma'} \nonumber \\
&+& s_\sigma \sum_{\tau'=n,p} \sum_{\sigma'=\up,\dw} s_{\sigma'} (b_1+c_1 + (b_3+c_3) I_{\tau} I_{\tau'}) K_{\tau'\sigma'} \nonumber \\
\label{SkmPotential}
\end{eqnarray}
is the Skyrme single-particle potential energy.

The electron contribution to the total energy density is
\begin{eqnarray}
{\cal E}_{elec}&=&\frac{eB}{(2\pi)^2}\sum_{N_e}\sum_{\sigma=\up,\dw} \nonumber \\
&\times& \int_{-\infty}^\infty dp_z
\varepsilon_{e\sigma}(p_z,N_e)f_{e\sigma}(\ee,T) \ ,
\label{eq:energden_elec}
\end{eqnarray}
where, $f_{e\sigma}(\ee,T)$ is the Fermi--Dirac distribution of electrons and
$\varepsilon_{e\sigma}(p_z,N_e)$ their single-particle energy which reads
\begin{equation}
\varepsilon_{e\sigma}(p_z,N_e)=\sqrt{m_e^2+2m_e\mu_BB(2N_e+1-g_e\frac{s_\sigma}{2})+p_z^2} \ ,
\label{eq:enere}
\end{equation}
with $m_e, N_e, \mu_B=5.78838 \times 10^{-15}$ MeV/G and $g_e=-2$ being, respectively, the mass, the Landau level, the Bohr magneton and the g-factor of the electron. The partial densities of spin up or spin down electrons and
the corresponding electron spin asymmetry density are
\begin{equation}
\rho_{e\sigma}=\frac{eB}{(2\pi)^2}\sum_{N_e}\int_{-\infty}^\infty dp_z f_{e\sigma}(\ee,T)
\label{eq:dense}
\end{equation}
and
\begin{equation}
W_e=\rho_{e\up}-\rho_{e\dw} \ ,
\label{eq:elec_asy}
\end{equation}
respectively. Note that, as in the case of protons, in Eqs.\ (\ref{eq:energden_elec}), (\ref{eq:dense}) and (\ref{eq:elec_asy}) the sum over the electron Landau levels $N_e$ runs from 0 up  to a maximum level obtained numerically as in the case of the maximum proton Landau level. For details, the reader is again referred to the appendix B of Ref.\ \cite{Bauer20}. 

The last remained element to be defined is the quantity $L_p$ (see Eq.\ (\ref{eq:energden})) of the explicit magnetic field contribution to the total energy density. This quantity is simply
\begin{equation}
L_p=\frac{eB}{(2\pi)^2}\sum_{N_p}N_p\sum_{\sigma=\up,\dw}\int_{-\infty}^\infty dp_z f_{p\sigma}(\ep,T) \ .
\label{eq:lp}
\end{equation}

Once we have the total energy density, the Helmhotz total free energy density, from which the chemical potentials of all the particle species can be evaluated, is easily obtained from the usual thermodynamical relation
\begin{equation}
{\cal F}={\cal E}-T{\cal S} \ ,
\label{eq:freeenergden}
\end{equation}
where ${\cal S}$ is total entropy density
\begin{equation}
{\cal S}={\cal S}_n+{\cal S}_p+{\cal S}_e \ ,
\label{eq:entropden}
\end{equation}
with ${\cal S}_n$, ${\cal S}_p$ and ${\cal S}_e$ the corresponding neutron, proton and electron contributions:
\begin{widetext}
\begin{eqnarray}
{\cal S}_n&=&-\sum_{\sigma=\up,\dw}\frac{1}{(2\pi)^3}\int d^3\vec p
\left[f_{n\sigma}\mbox{ln}(f_{n\sigma})+(1-f_{n\sigma})\mbox{ln}(1-f_{n\sigma})
\right] \ , \nonumber \\
{\cal S}_p&=&-\sum_{\sigma=\up,\dw}\frac{eB}{(2\pi)^2}\sum_{N_p}\int_{-\infty}^{\infty}dp_z
\left[f_{p\sigma}\mbox{ln}(f_{p\sigma})+(1-f_{p\sigma})\mbox{ln}(1-f_{p\sigma})
\right] \ , \nonumber \\
{\cal S}_e&=&-\sum_{\sigma=\up,\dw}\frac{eB}{(2\pi)^2}\sum_{N_e}\int_{-\infty}^{\infty}dp_z
\left[f_{e\sigma}\mbox{ln}(f_{e\sigma})+(1-f_{e\sigma})\mbox{ln}(1-f_{e\sigma})
\right] \ ,
\label{eq:entropdenpart}
\end{eqnarray}
\end{widetext}
where we have omitted the explicit dependencies of the Fermi--Dirac distributions to simplify the notation.

Once ${\cal F}$ is known one can obtain the pressure of the system simply as
\begin{equation}
P=\rho\left(\frac{\partial {\cal F}}{\partial \rho}\right)_{T,B}-{\cal F} \ ,
\label{eq:pressure}
\end{equation}
from which is possible to determine the isothermal compressibility
\begin{equation}
{\cal K}=\left[\rho\left(\frac{\partial P}{\partial \rho}\right)_{T,B} \right]^{-1} \ .
\label{eq:isothcom}
\end{equation}

%%%%%%%%%%%%%%%%%%%%%%%%%%%%%%%%%%%%%%%%%%%%%%%%%%%%%%%%%%%%%%%%%%%
\section{Results and Discussion}
\label{sec:res}

In the following we discuss the properties of hot and dense neutron star matter under the presence of strong magnetic fields. Results are presented for densities up to 0.4 fm$^{-3}$, temperatures $T=5$, $15$ and $30$ MeV, and the magnetic fields strengths $B=10^{16}$, $10^{17}$ and $10^{18}$ G.

%%%%%%%%%%%%%%%%%%%%%%%%%%%%%%%%%%%%%%%%%%%%%%%%%%%%%%%%
\begin{figure*}[t!]
\centering
\includegraphics[width=1.5 \columnwidth]{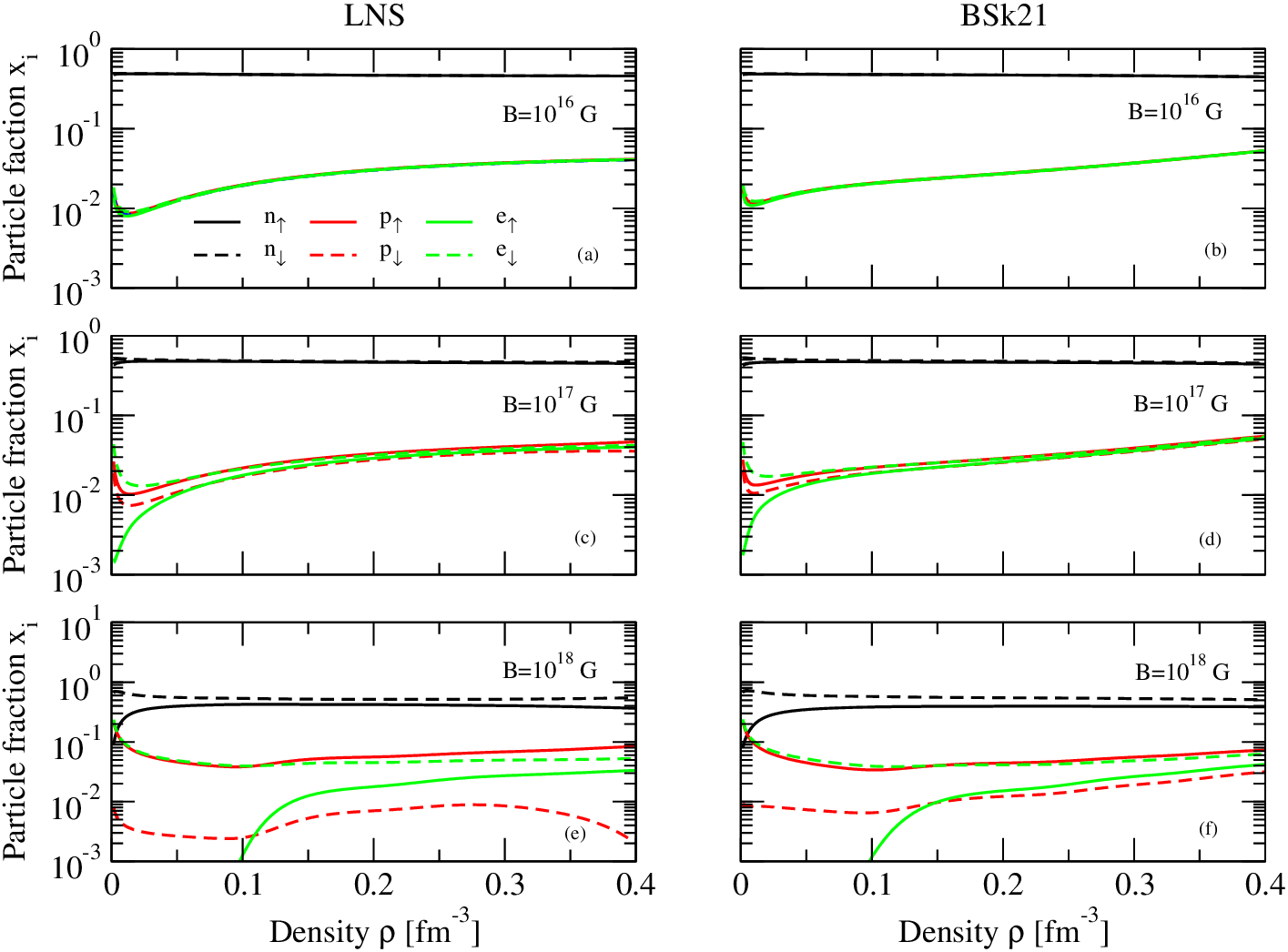}
\caption{(Color online) Particle fractions $x_i\equiv \rho_i/\rho$ ($i=n_\up, n_\dw, p_\up, p_\dw, e_\up, e_\dw$) of $\beta$-stable matter at $T=5$ MeV for the magnetic field strengths, $10^{16}, 10^{17}$ and $10^{18}$ G for the two interactions considered, LNS (panels (a), (c) and (e)) and BSk21 (panels (b), (d) and (f)).}
\label{fig:fig1}
\end{figure*}
%%%%%%%%%%%%%%%%%%%%%%%%%%%%%%%%%%%%%%%%%%%%%%%%%%%%%%%%

We start by showing in Fig.\ \ref{fig:fig1} the fractions of neutrons, protons and electrons with spin up and down ($x_i\equiv \rho_i/\rho$,  $i=n_\up, n_\dw, p_\up, p_\dw, e_\up, e_\dw$) in $\beta$-stable matter, obtained by solving Eqs.\ (\ref{baryon}), (\ref{charge}) and (\ref{chempot2}), at $T=5$ MeV for the three magnetic field strengths just mentioned and for the two interactions considered, LNS (panels (a), (c) and (e)), and BSk21 (panels (b), (d) and (f)). As it is seem in the figure, a magnetic field of strength $10^{16}$ G induces only a extremely low polarization of the spins of the different components of neutron star matter at very low densities, and fields of the order of at least $10^{17}$ G are needed to see appreciable differences in the fractions of neutrons, protons and electrons with opposite spin projections. We note also that whereas the fraction of protons with spin up ({\it i.e.,} oriented parallel to the magnetic field) is larger than the fraction of protons with spin down the opposite is observed in the case of neutrons and electrons. This is simply a consequence of the fact that the proton g-factor is positive while the neutron and electron ones are negative. Due to this, spin up (down) protons (neutrons and electrons) have lower energy than spin down (up) protons (neutrons and electrons) (see Eqs.\ (\ref{eq:enern}),  (\ref{eq:enerp}) and (\ref{eq:enere})). Consequently, the configurations of the system with fractions of spin up protons larger than spin down protons, and fractions of spin down neutrons and electrons larger than spin up neutrons and electrons, have less energy and, therefore, are physically favorable. We observe that although in the case of the BSk21 interaction, for each particle species $i$, the difference in the fractions between the spin up and spin down component decreases with increasing density, in the case of the LNS an increase of this difference is observed for neutrons and protons for densities $\rho \gtrsim 0.3$ fm$^{-3}$. This is a consequence of the appearance of a ferromagnetic instability predicted by the LNS model at high densities, instability that is corrected in the case of the BSk21 force as we mentioned before.

%%%%%%%%%%%%%%%%%%%%%%%%%%%%%%%%%%%%%%%%%%%%%%%%%%%%%%%%
\begin{figure}[b!]
\centering
\includegraphics[width=1 \columnwidth]{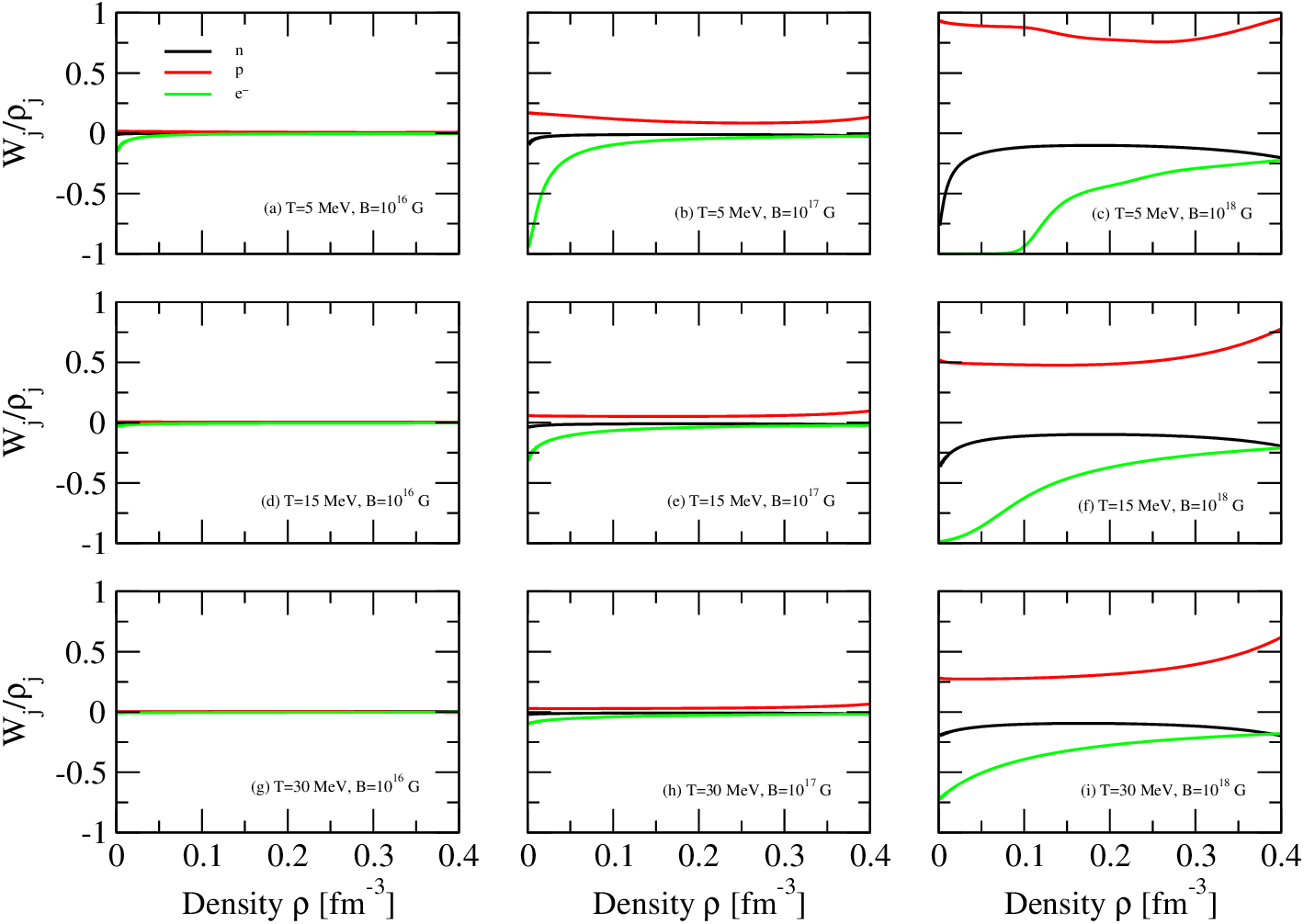}
\caption{(Color online) Spin asymmetry $W_j/\rho_j$ of each particle species for neutron star matter at several temperatures and magnetic field strengths for the LNS interaction.}
\label{fig:fig2}
\end{figure}
%%%%%%%%%%%%%%%%%%%%%%%%%%%%%%%%%%%%%%%%%%%%%%%%%%%%%%%%
%%%%%%%%%%%%%%%%%%%%%%%%%%%%%%%%%%%%%%%%%%%%%%%%%%%%%%%%
\begin{figure}[b!]
\centering
\includegraphics[width=1 \columnwidth]{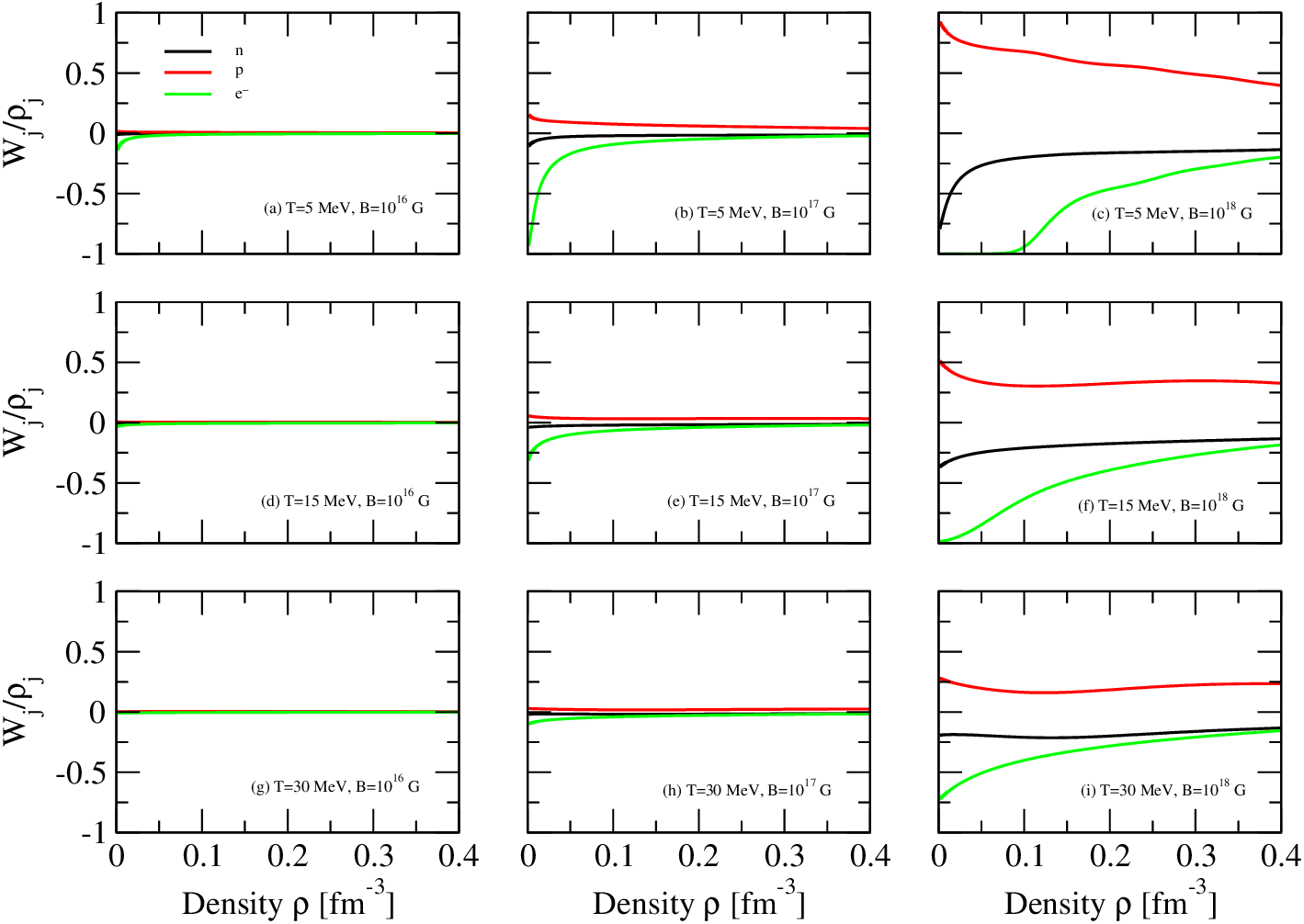}
\caption{(Color online) Spin asymmetry $W_j/\rho_j$ of each particle species for neutron star matter at several temperatures and magnetic field strengths
for the BSk21 interaction.}
\label{fig:fig3}
\end{figure}
%%%%%%%%%%%%%%%%%%%%%%%%%%%%%%%%%%%%%%%%%%%%%%%%%%%%%%%%
%%%%%%%%%%%%%%%%%%%%%%%%%%%%%%%%%%%%%%%%%%%%%%%%%%%%%%%%
\begin{figure*}[t!]
\centering
\includegraphics[width=1.5 \columnwidth]{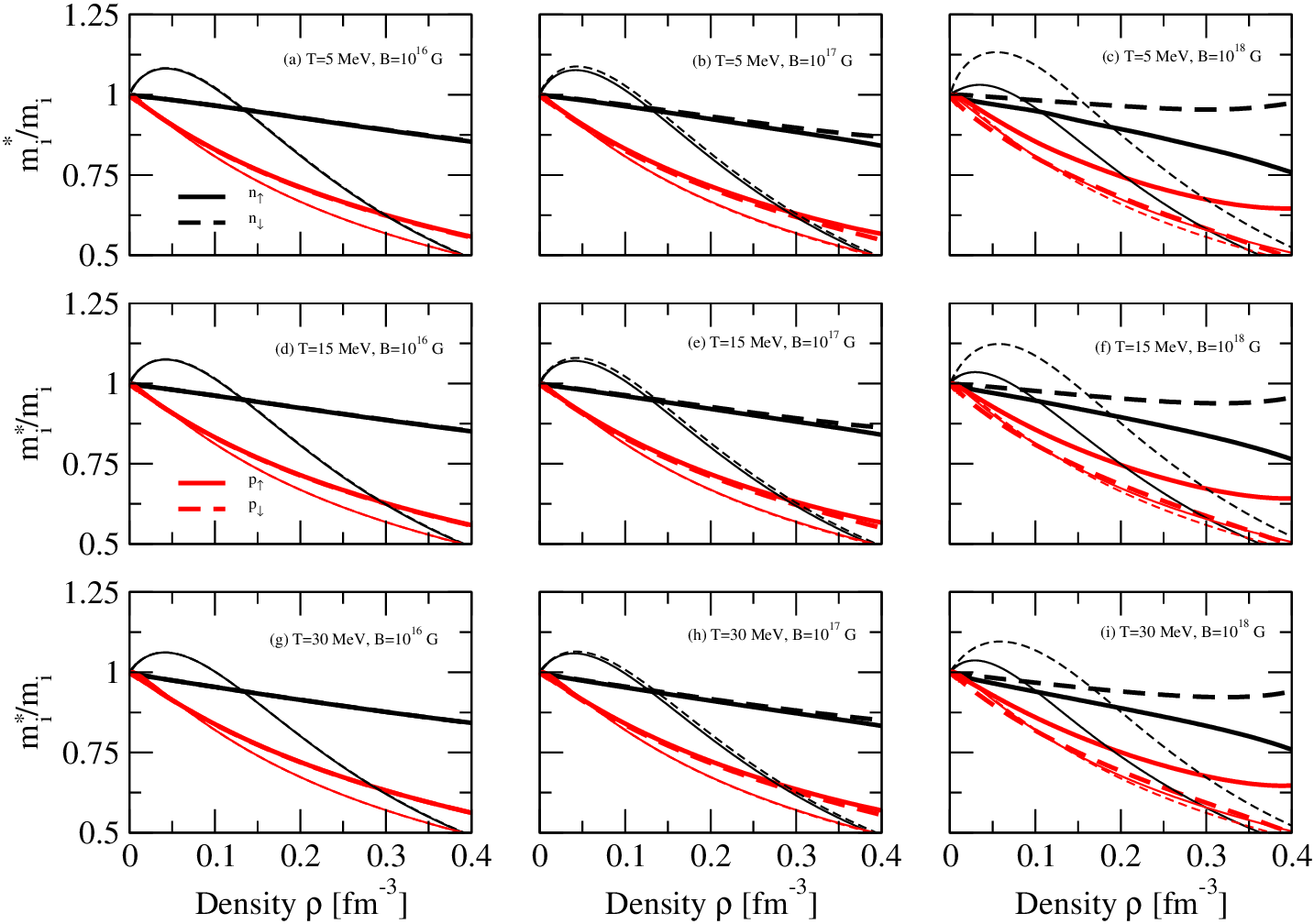}
\caption{(Color online) Neutron and proton effective masses predicted by the LNS (thick lines) and the BSk21 (thin lines) interactions as function of the density for the three temperatures considered in this work.}
\label{fig:fig4}
\end{figure*}
%%%%%%%%%%%%%%%%%%%%%%%%%%%%%%%%%%%%%%%%%%%%%%%%%%%%%%%%

To understand better the state of spin polarization of the system, we show now the spin asymmetry, defined as the ratio $W_j/\rho_j=(\rho_{j\up}-\rho_{j\dw})/\rho_j$ with $\rho_j=\rho_{j\up}+\rho_{j\dw}$ ($j=n,p,e$), of each particle species for the three temperatures and the three magnetic fields considered in this work. Results for the LNS and the BSk21 models are presented in Fig.\ \ref{fig:fig2} and Fig.\ \ref{fig:fig3}, respectively. We note first that the value $W_j/\rho_j=0$ corresponds to the case in which the species $j$ is unpolarized, whereas $W_j/\rho_j=\pm 1$ means that this species  is totally polarized, {\it i.e.,} all its spins are aligned along the same direction, parallel ($W_j/\rho_j=1$) or antiparallel ($W_j/\rho_j=-1$) to the one defined by the magnetic field. Partially polarized configurations of a species $j$ correspond to values of $W_j/\rho_j$ between $-1$ and $1$. To begin with, we observe in both figures that a magnetic field of $10^{16}$ G has almost no effect on the spin asymmetry of the different particles, being the system essentially in a global unpolarized state. Just for the lowest temperature $T=5$ MeV and at very low densities this field induces a very tiny polarization of the particle spins, especially on the electron ones. Only magnetic fields with a strength $B\geq10^{17}$ G are able to change the spin polarization of the system from the unpolarized state to a partially polarized one. As the sign of the spin asymmetry of each particle indicates, while protons have their majority of their spins oriented parallel to the magnetic field ($W_p/\rho_p > 0$), neutron and electron spins are mostly aligned in 
the opposite direction ($W_n/\rho_n<0$, $W_e/\rho_e<0$). This, as it was discussed above, is because system configurations with larger fractions of spin up protons and spin down neutrons and electrons are energetically favored over those with larger fractions of spin down protons and spin up neutrons and electrons. We note that although the spin asymmetry of the protons (neutrons) decreases (increases) always with density for the BSk21 interaction, indicating the opposition of this nuclear interaction to the spin polarization induced by the magnetic field, this is not the case when the LNS one is used.  In this case, we observe that $W_p/\rho_p$ ($W_n/\rho_n$) decreases (increases) of  up to a density $\rho \sim 0.2$ fm$^{-3}$ for $B=10^{17}$ G ($\rho \sim 0.3$ fm$^{-3}$ for $B=10^{18}$ G) and then it increases (decreases). This behavior is again just a consequence of the ferromagnetic instability predicted by the LNS force. Regarding the electrons, we see that their spin asymmetry always increases monotonously with density, reaching asymptotically their unpolarized state ($W_e/\rho_e=0$) at high densities. We observe also that protons and electrons are more polarized than neutrons in all the range of densities explored, being the lower degree of polarization of the neutrons due to its weak anomalous magnetic moment. Note finally, that the spin asymmetry of the three species decreases (in absolute value) when increasing the temperature. This is expected since increasing temperature increases the entropy of the system and, consequently, its disorder. The number of spin up and spin down particles becomes more and more similar and, therefore, the system and becomes less polarized.

\begin{figure}[t!]
\centering
\includegraphics[width=1 \columnwidth]{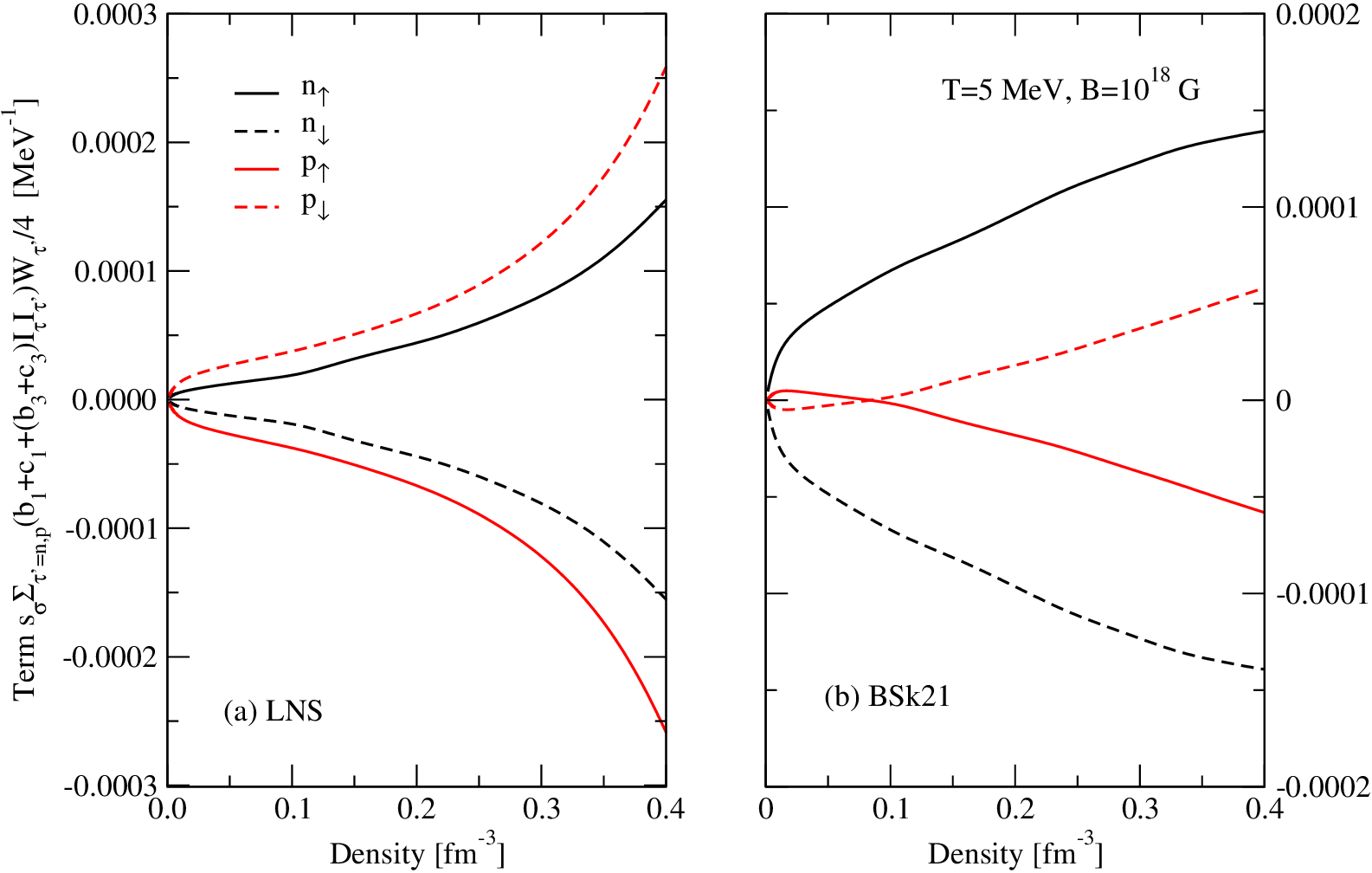}
\caption{(Color online) Density dependence of the term $\frac{s_\sigma}{4}\sum_{\tau'=n,p}(b_1+c_1+(b_3+c_3)I_\tau I_{\tau'})W_{\tau'}$ of Eq.\ (\ref{eq:effmass}) for the LNS (panel (a)) and BSk21 (panel (b)) interactions for neutron star matter at $T=5$ MeV in the presence of a magnetic field of $10^{18}$ G.}
\label{fig:fig5}
\end{figure}

In Fig.\ \ref{fig:fig4} we show now the effect of the magnetic field on the neutron and proton effective masses predicted by the LNS (thick lines) and the BSk21 (thin lines) interactions for the three temperatures considered in this work. First, we observe that a magnetic field of $10^{16}$ G is too low to have any effect on the effective mass of both neutrons and protons with different spin projections because, as
shown in panels (a), (d) and (g) of Figs.\  \ref{fig:fig2} and \ref{fig:fig3}, it has almost no effect on their spin asymmetry and, consequently, their effective masses are essentially spin independent for this field. Second, we note that
the increase of  $m^*_{\tau\sigma}/m_\tau$ above one seen in the case of the BSk21 interaction is a consequence of the density dependence of the effective mass predicted by this force which goes as $(a+b\rho+c\rho^{1+\beta}+d\rho^{1+\gamma})^{-1}$ (see Eq.\ (\ref{eq:effmass})), whereas the LNS interaction, for which the coefficients $c_0, \cdot\cdot\cdot, c_3$ are zero, predicts
$m^*_{\tau\sigma} \sim (\tilde a+\tilde b\rho)^{-1}$. Finally, we notice for both interactions that the effective mass of neutrons and protons is in general larger for the more abundant of their spin projection component, respectively, spin down neutrons and spin up protons. Note, however, that the BSk21 interaction in the case of protons predicts $m^*_{p_\dw}>m^*_{p_\up}$ for densities below $\sim 0.1$ fm$^{-3}$. The splitting of the spin up and spin down nucleon effective masses can be understood by looking at the term $\frac{s_\sigma}{4}\sum_{\tau'=n,p}(b_1+c_1+(b_3+c_3)I_\tau I_{\tau'})W_{\tau'}$ of Eq.\ (\ref{eq:effmass}). To facilitate the analysis of the role of this term on the effective masses, in Fig.\ \ref{fig:fig5} we show its density dependence for the two interactions considered for neutron star matter at $T=5$ MeV in the presence of  a magnetic field of $10^{18}$ G. Let us consider first the case of the LNS interaction. As it is seen in panel (a), for this model this term is always positive for spin up neutrons and spin down protons while it is negative for spin down neutrons and spin up protons. Therefore, is it clear from the definition of the effective mass given in Eq.\  (\ref{eq:effmass}) that, for the LNS interaction, the contribution of this term leads to values of $m^*_{n_\dw}$ and $m^*_{p_\up}$ always larger than those of $m^*_{n_\up}$ and $m^*_{p_\dw}$, respectively.  For the BSk21 interaction (panel (b)) we observe that also in this case this term is always positive for spin up neutrons and negative for spin down neutrons and, therefore, $m^*_{n_\dw}>m^*_{n_\up}$ in the whole range of densities explored. On the other hand, for densities below $\sim 0.1$ fm$^{-3}$, this term is positive (negative) for spin up (down) protons and   vice versa for densities above this value. Therefore, as it is seen in Fig.\ \ref{fig:fig4}, this interaction predicts $m^*_{p_\dw}>m^*_{p\up}$ for $\rho < 0.1$ fm$^{-3}$ and $m^*_{p_\up}> m^*_{p_\dw}$ for $\rho > 0.1$ fm$^{-3}$. Similar conclusions can be drawn from the analysis of results obtained for other temperatures and magnetic fields.

%%%%%%%%%%%%%%%%%%%%%%%%%%%%%%%%%%%%%%%%%%%%%%%%%%%%%%%%
%%%%%%%%%%%%%%%%%%%%%%%%%%%%%%%%%%%%%%%%%%%%%%%%%%%%%%%%

\begin{figure}[t!]
\centering
\includegraphics[width=1 \columnwidth]{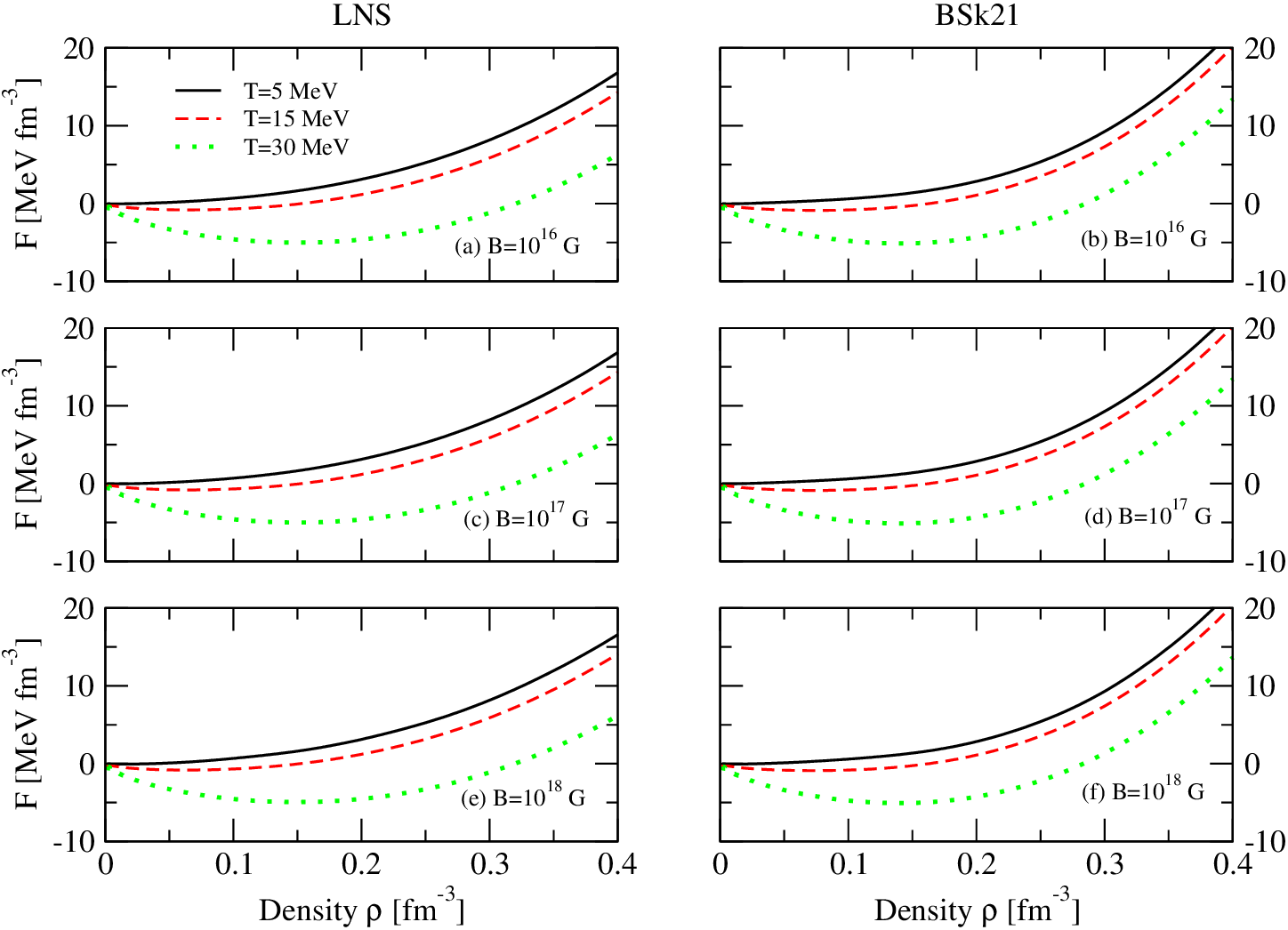}
\caption{(Color online) Helmhotz total free energy density as a function of the density for the two interactions and the different temperatures and magnetic fields considered in this work.}
\label{fig:fig6}
\end{figure}

Let us finish this section by analyzing the bulk thermodynamical properties of the system. We show first in Fig.\ \ref{fig:fig6} the Helmhotz total free energy density as a function of the density for the two interactions and the different temperatures and magnetic fields considered in this work. As it can be seen the effect of the magnetic field seems to be almost negligible. The reason is mainly due to the low value of the nuclear magneton which makes the interaction of neutrons and protons with the magnetic field too mild for the values of $B$ considered. Consequently, the contribution to the energy density from the interaction of nucleons with the field (last term of Eq.\ (\ref{eq:energden})) is too small, and that from the nuclear interaction ${\cal E}_{nuc}$ (Eq.\ (\ref{eq:energden_nuc})) depends also very little on it. In addition, also the electron contribution ${\cal E}_{elec}$ (Eq.\ (\ref{eq:energden_elec})) to the total energy density depends very weakly on the field, increasing slightly when increasing $B$. The reason, in this case, should not be attributed to the value of the Bohr magneton, three orders of magnitude larger than the nuclear one, but rather to the fact that the total electron fraction is very low (see Fig.\ \ref{fig:fig1}). For illustration we show, for the two model interactions considered, in Tabs.\ \ref{tab:tab1} (LNS) and \ref{tab:tab2} (BSk21) all the contributions to the Helmholtz total free energy density for three representative densities $\rho=0.08$ fm$^{-3}$, $\rho=0.16$ fm$^{-3}$ and $\rho=0.32$ fm$^{-3}$, a temperature $T$ of 5 MeV, and the magnetic fields $B=10^{16}$ G and $B=10^{18}$ G. Note that also the neutron, proton and electron contributions to the total entropy density depend very little on the magnetic field.

\begin{figure}[t!]
\centering
\includegraphics[width=1 \columnwidth]{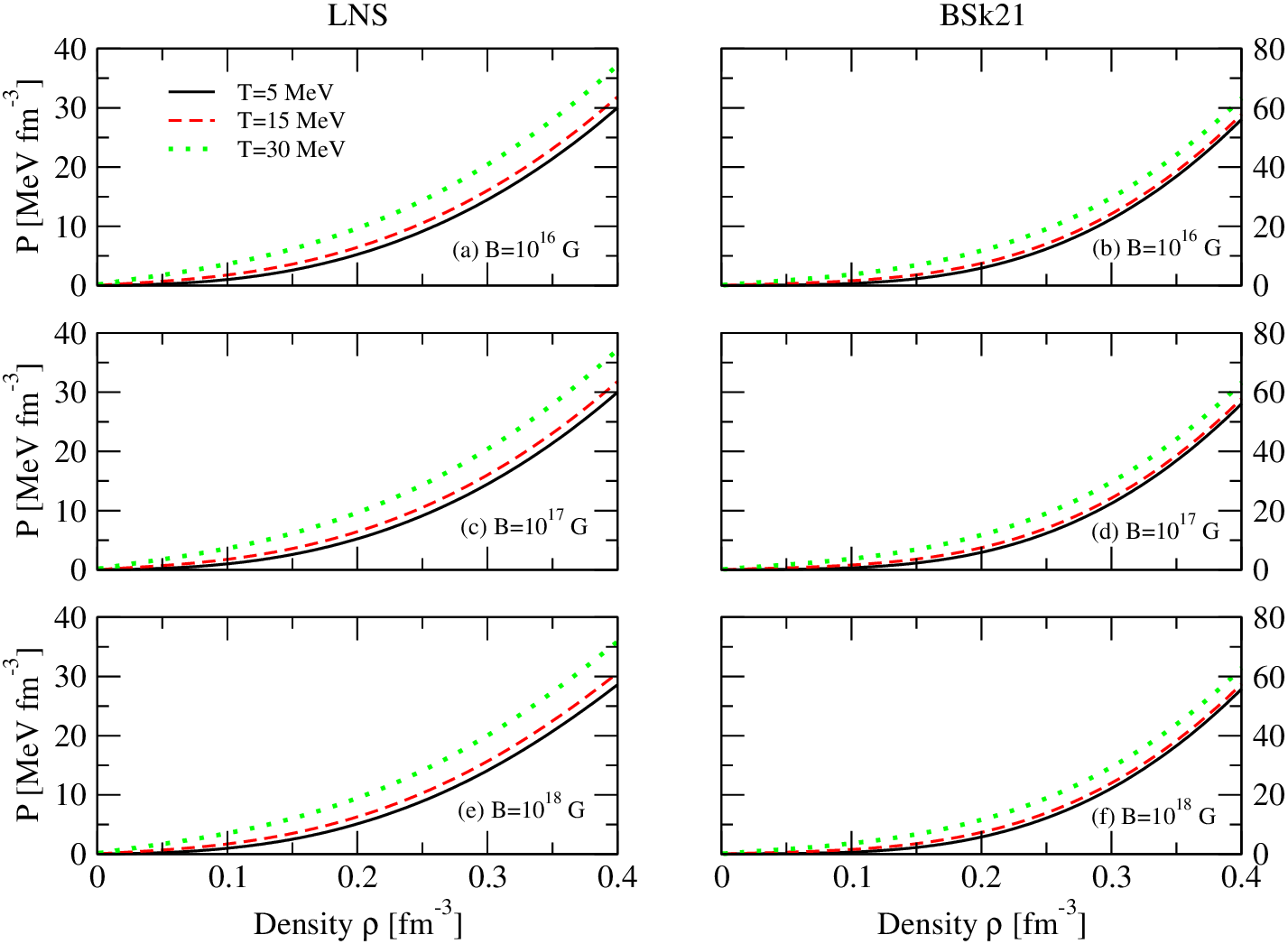}
\caption{(Color online) Pressure of the system as a function of the density for the two interactions and the different temperatures and magnetic fields considered in this work.}
\label{fig:fig7}
\end{figure}

\begin{figure}[t!]
\centering
\includegraphics[width=1 \columnwidth]{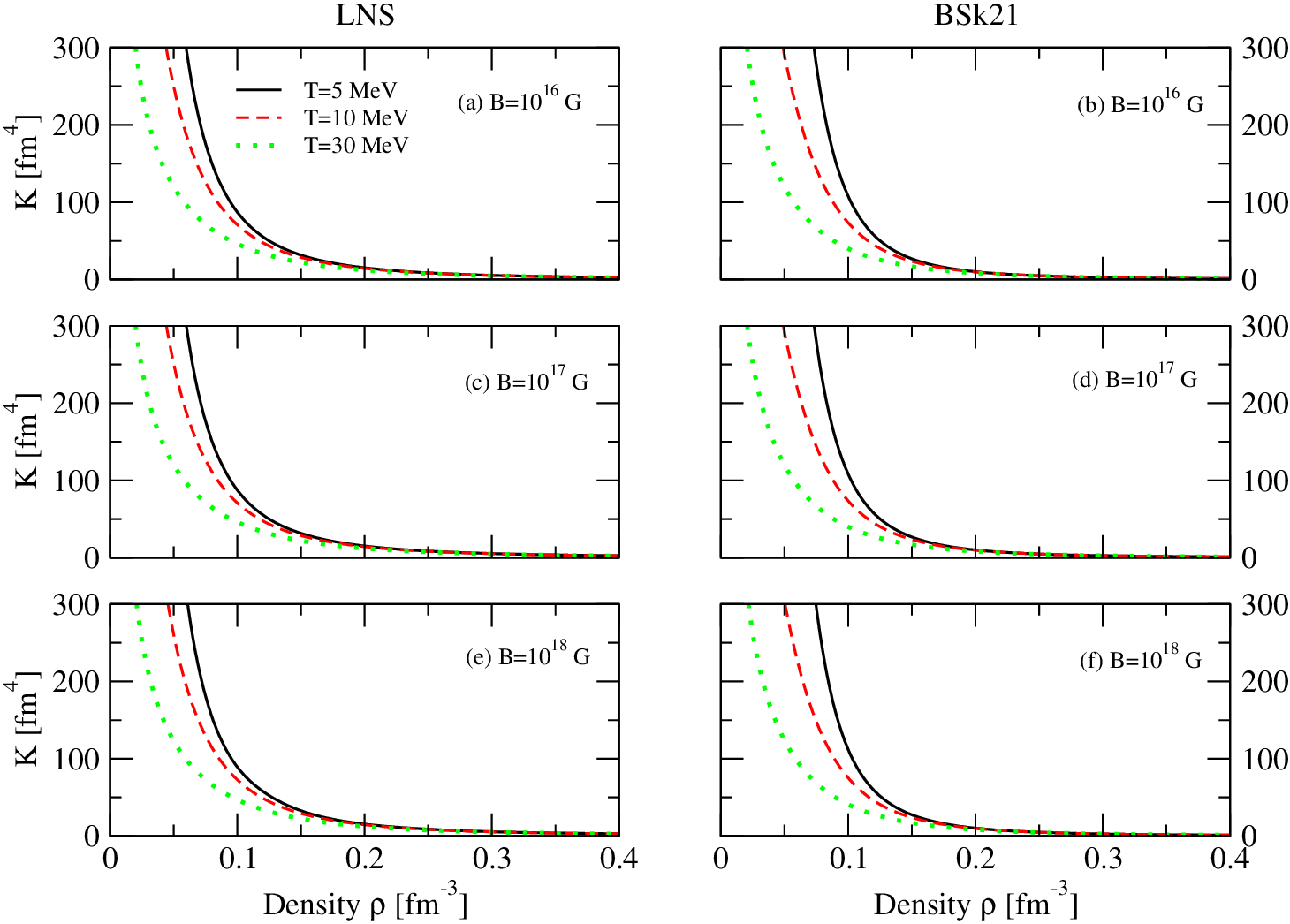}
\caption{(Color online) Isothermal compressibility as a function of the density for the two interactions and the different temperatures and magnetic fields considered in this work.}
\label{fig:fig8}
\end{figure}

Finally, in Figs.\ \ref{fig:fig7} and \ref{fig:fig8} we show, respectively, the pressure and the isothermal compressibility of the system as a function of the density for the two interactions and the different temperatures and magnetic fields considered in this work. The pressure, as it is required by the stability conditions, increases monotonically with the density and is larger for larger values of the temperature. Note that, as expected from our previous analysis of the Helmhotz total free energy density, both the pressure and the isothermal compressibility present also a very mild dependence on the magnetic field which is almost imperceptible in the figures. As it is seen, the isothermal compressibility decreases monotonously with density from relatively high values at low densities, and it becomes very small for $\rho \gtrsim 0.3$ fm$^{-3}$, showing that from this density on (for these two particular interactions) highly magnetized neutron star matter can be considered an almost incompressible system.

%%%%%%%%%%%%%%%%%%%%%%%%%%%%%%%%%%%%%%%%%%%%%%%%%%%%%%%%
%%%%%%%%%%%%%%%%%%%%%%%%%%%%%%%%%%%%%%%%%%%%%%%%%%%%%%%%

%%%%%%%%%%%%%%%%%%%%%%%%%%%%%%%%%%%%%%%%%%%%%%%%%%%%%%%%
%%%%%%%%%%%%%%%%%%%%%%%%%%%%%%%%%%%%%%%%%%%%%%%%%%%%%%%%

%%%%%%%%%%%%%%%%%%%%%%%%%%%%%%%%%%%%%%%%%%%%%%%%%%%%%%%%
%%%%%%%%%%%%%%%%%%%%%%%%%%%%%%%%%%%%%%%%%%%%%%%%%%%%%%%%

%%%%%%%%%%%%%%%%%%%%%%%%%%%%%%%%%%%%%%%%%%%%%%%%%%
\begin{table*}[t]
\begin{center}
\begin{tabular}{c|cccccccc}
\hline
\hline
$B$ &$\rho$ & ${\cal E}_{nucl}$ & ${\cal E}_{elec}$ & $\mu_NB\left(2L_p +\rho_p-\frac{g_p}{2}W_p+\frac{g_n}{2}W_n\right)$ &
${\cal S}_pT$ & ${\cal S}_nT$ & ${\cal S}_eT$ & ${\cal F}$   \\
\hline
& $0.08$ & $0.625$ & $0.088$ & $0.014$ & $0.042$& $0.253$ & $0.008$ & $0.424$   \\
$10^{16}$& $0.16$ & $1.836$ & $0.402$ & $0.056$ & $0.086$ & $0.310$ & $0.016$ & $1.882$   \\
& $0.32$ & $8.264$ & $1.630$ & $0.237$ & $0.141$ & $0.365$ & $0.033$ & $9.592$   \\
\hline
& $0.08$ & $0.606$ & $0.144$ & $-0.064$ & $0.043$ & $0.251$ & $0.004$ & $0.388$   \\
$10^{18}$ & $0.16$ & $1.783$ & $0.563$ & $-0.084$ & $0.079$ & $0.310$ & $0.018$ & $1.855$   \\
& $0.32$ & $8.259$ & $1.939$ & $-0.108$ & $0.123$ & $0.367$ & $0.031$& $9.569$   \\
\hline
\hline
\end{tabular}
\end{center}
\caption{Separate contributions to the Helmhotz total free energy density for densities $\rho=0.08$ fm$^{-3}$, $\rho=0.16$ fm$^{-3}$ and $\rho=0.32$ fm$^{-3}$, a temperature $T=5$ MeV, and the magnetic fields $B=10^{16}$ G and $B=10^{18}$ G for the LNS interaction. The Helmhotz total free energy density is shown in the last column. Units are given in MeV fm$^{-3}$.}
\label{tab:tab1}
\end{table*}
%%%%%%%%%%%%%%%%%%%%%%%%%%%%%%%%%%%%%%%%%%%%%%%%%%%%%%%%%%%%%%%%%%%%

%%%%%%%%%%%%%%%%%%%%%%%%%%%%%%%%%%%%%%%%%%%%%%%%%%
\begin{table*}[t]
\begin{center}
\begin{tabular}{c|cccccccc}
\hline
\hline
$B$ &$\rho$ & ${\cal E}_{nucl}$ & ${\cal E}_{elec}$ & $\mu_NB\left(2L_p +\rho_p-\frac{g_p}{2}W_p+\frac{g_n}{2}W_n\right)$ &
${\cal S}_pT$ & ${\cal S}_nT$ & ${\cal S}_eT$ & ${\cal F}$   \\
\hline
& $0.08$ & $0.612$ & $0.104$ & $0.017$ & $0.046$& $0.272$ & $0.008$ & $0.407$   \\
$10^{16}$& $0.16$ & $1.590$ & $0.362$ & $0.050$ & $0.082$ & $0.296$ & $0.016$ & $1.608$   \\
& $0.32$ & $9.700$ & $1.720$ & $0.257$ & $0.139$ & $0.245$ & $0.035$ & $11.258$   \\
\hline
& $0.08$ & $0.635$ & $0.151$ & $-0.100$ & $0.046$ & $0.271$ & $0.004$ & $0.365$   \\
$10^{18}$ & $0.16$ & $1.623$ & $0.489$ & $-0.140$ & $0.079$ & $0.297$ & $0.019$ & $1.577$   \\
& $0.32$ & $9.880$ & $1.964$ & $-0.087$ & $0.134$ & $0.247$ & $0.032$ & $11.344$   \\
\hline
\hline
\end{tabular}
\end{center}
\caption{As Tab.\ \ref{tab:tab1} for the BSk21 interaction.}
\label{tab:tab2}
\end{table*}
%%%%%%%%%%%%%%%%%%%%%%%%%%%%%%%%%%%%%%%%%%%%%%%%%%%%%%%%%%%%%%%%%%%%

%%%%%%%%%%%%%%%%%%%%%%%%%%%%%%%%%%%%%%%%%%%%%%%%%%%%%%%%%%%%%%%%%%%
\section{Summary and conclusions}
\label{sec:suc}
In the present work we have studied the properties of hot and dense neutron star matter under the presence of  strong magnetic fields using two Skyrme interactions, namely the LNS interaction developed by Cao {\it et al.,} \cite{Cao06} and the interaction BSk21 \cite{Goriely10} of the Brussels--Montreal group. In particular, we have constructed the equation of state of the system and analyze its composition for a range of densities, temperatures and magnetic field intensities of interest for the study of supernova and proto-neutron star matter, with a particular interest on the degree of
spin-polarization of the different components. Our results show that in order to see appreciable differences in the fractions of neutrons, protons and electrons with opposite spin projections the intensity of the magnetic field should be at least of the order of $10^{17}$ G. They also show that system configurations with larger fractions of spin up protons and spin down neutrons and electrons are energetically favored over those with larger fractions of spin down protons and spin up neutrons and electrons. We have also studied the effect of the magnetic field
on the neutron and proton effective masses finding that, for the two interactions considered, that the effective mass of neutrons and protons is in general larger for the more abundant of their spin projection component, respectively, spin down neutrons and spin up protons. Finally, we have determined the bulk thermodynamical properties of the system, finding that the effect of the magnetic field on the Helmhotz total free energy density, pressure and isothermal compressibility of the system is almost negligible due to the low value of the nuclear magneton which makes the
interaction of neutrons and protons with the magnetic field too mild for the values of B considered. 

%%%%%%%%%%%%%%%%%%%%%%%%%%%%%%%%%%%%%%%%%%%%%%%%%%%%%%%%%%%%%%%%%%%

\section*{Acknowledgements}

I.V. thanks the support of the European Union’s Horizon 2020 research and innovation programme under Grant Agreement No. 824093.

%%%%%%%%%%%%%%%%%%%%%%%%%%%%%%%%%%%%%%%%%%%%%%%%%%%%%%%%%%%%%%%%%%%%%%%%%%%%%%%%%%%%

%%%%%%%%%%%%%%%%%%%%%%%%%%%%%%%%%%%%%%%%%%%%%%%%%%%%%%%%%%%%%%%%%%%%%%%%%%%%%%%%%%%%

%%%%%%%%%%%%%%%%%%%%%%%%%%%%%%%%%%%%%%%%%%%%%%%%%%%%%%%%%%%%%%%%%%%%%%%%%%%%%%%%%%%%

\end{document}